\documentclass[letterpaper]{article}
\usepackage{iccc}

\usepackage{times}
\usepackage{helvet}
\usepackage{courier}
\usepackage{csquotes}
\usepackage{graphicx}

\title{Collaborative Storytelling with Human Actors and AI Narrators \\
Paper type: Event Report}
\author{Boyd Branch\\
University of Kent\\
United Kingdom\\
boyd@improvmedialab.com\\
\And
Piotr Mirowski\\
Improbotics\\
United Kingdom\\
improbotics.org\\
\And
Kory Mathewson\\
Improbotics\\
Canada\\
improbotics.org\\
}

\begin{document} 
\maketitle

\begin{abstract}
\begin{quote}
Large language models can be used for collaborative storytelling. In this work we report on using GPT-3 \cite{brown2020language} to co-narrate stories. The AI system must track plot progression and character arcs while the human actors perform scenes. This event report details how a novel conversational agent was employed as creative partner with a team of professional improvisers to explore long-form spontaneous story narration in front of a live public audience. We introduced novel constraints on our language model to produce longer narrative text and tested the model in rehearsals with a team of professional improvisers. We then field tested the model with two live performances for public audiences as part of a live theatre festival in Europe. We surveyed audience members after each performance as well as performers to evaluate how well the AI performed in its role as narrator. Audiences and performers responded positively to AI narration and indicated preference for AI narration over AI characters within a scene. Performers also responded positively to AI narration and expressed enthusiasm for the creative and meaningful novel narrative directions introduced to the scenes. Our findings support improvisational theatre as a useful test-bed to explore how different language models can collaborate with humans in a variety of social contexts.
\end{quote}
\end{abstract}

\section{Improv Theatre and AI}
Improvisational theatre is increasingly being used as an environment and platform for testing and exploring the creative potential of computationally creative systems \cite{bruce2000robot,baumer2010analysis,o2011knowledge,jacob2019improvisational} and of artificial intelligence language models in particular \cite{martin2016improvisational,mathewson2017improvised,mathewson2018improbotics,cho2020grounding}. 
Turing test inspired experiments focus on evaluating how well language models can perform natural-sounding human language \cite{turing1951can}. Conversely, improvisational theatre is uniquely positioned to explore the collaborative potential of language models. 
Collaborative storytelling includes both on-stage performance (e.g. improvised theatre) and off-stage games (e.g. table-top role-playing games, card games).
Collaborative storytelling in collaboration with artificial agents has been studied previously \cite{perlin1996improv,hayes1996improvisational,riedl2006believable,magerko2011employing}, most often in the context of virtual environments where the human players interact with digital avatars, with the exception of plot generation tools like \emph{dAIrector} \cite{eger2018dairector} informing live improv on stage. 
Recent advances in large language models enable richer text-based interaction between human and AI players \cite{nichols2020collaborative}, as illustrated by the success of online role-playing game \emph{AI Dungeon}\footnote{\small{https://play.aidungeon.io/}}. 
Our case study pulls away from the virtual world and situates AI and human collaborators together on-stage. This shared narrative is then interpreted by live human actors, expressing the full range of emotional, physical and verbal human creativity.

Improvised theatre explores how interesting narratives can emerge from establishing rules for simple social dynamics and rhetorical conventions. In contrast to scripted theatre, improv is built from spontaneity \cite{spolin1963improvisation}. Improvisers are trained to disengage executive cognition in order to allow their automatic responses to guide and justify a given and emerging social context \cite{johnstone1979}. Narratives emerge by assuming the presence of meaning. The performer only needs to accept \emph{offers}: what is said and done on stage. There are no `wrong' things a performer can say to invalidate the emerging narrative.
For meaning to emerge for an audience however, each novel narrative statement must be followed up with some degree of agreement and justification. 
An improvisational scene is ultimately judged on the degree to which novel statements can be integrated back into the previous given circumstances. That process is called \emph{justification} and is synonymous with an ongoing adaptation, by the actors--thrown out of their comfort zone--to the changing dynamics of an improvised narration. This practice is what makes improv theatre such a useful platform to explore the creative capacity of artificial intelligence. For the AI to perform `well' it cannot simply introduce novel narrative subjects, but must also be able to adapt to the emerging given circumstances, akin to the desiderata of AI systems capable of generalising to unseen data.

Previously, \cite{mathewson2017improvised,mathewson2018improbotics,cho2020grounding} explored how an AI trained on movie or improv dialogue could generate interesting narratives as a \emph{performer} within an improvised scene, and demonstrated that conversational agents built using recurrent neural networks or transformers, e.g., GPT-2 \cite{radford2019language}, could indeed move a given narrative forward when human agents were operating to accept and justify the statements. In their setup, the AI only functioned as a character in a given scene. In our study, we examine how AI performs in the role of the narrator.


\section{Methods}


\begin{figure}[t]
    \centering
    \includegraphics[width=0.49\columnwidth]{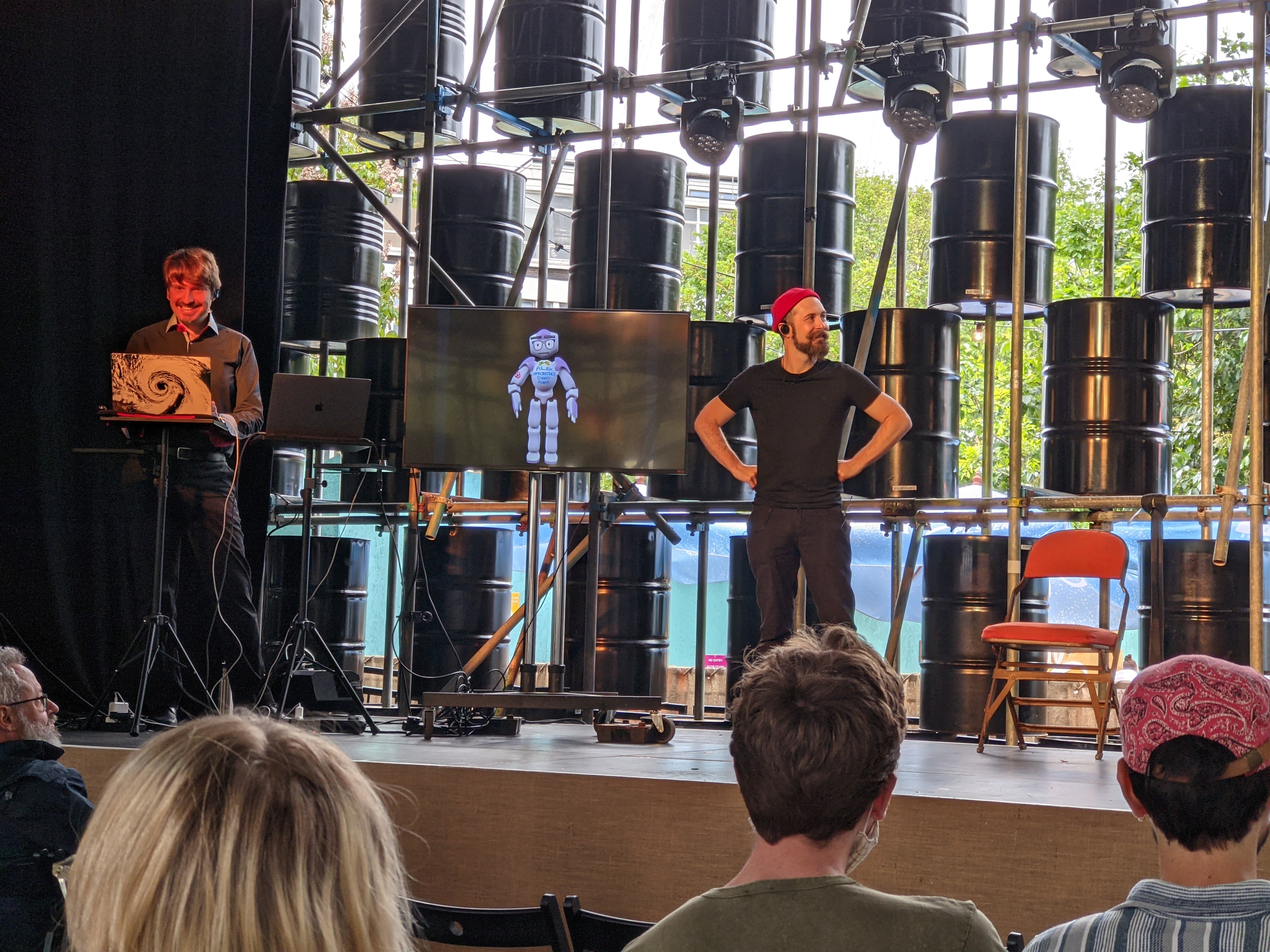}
    \includegraphics[width=0.49\columnwidth]{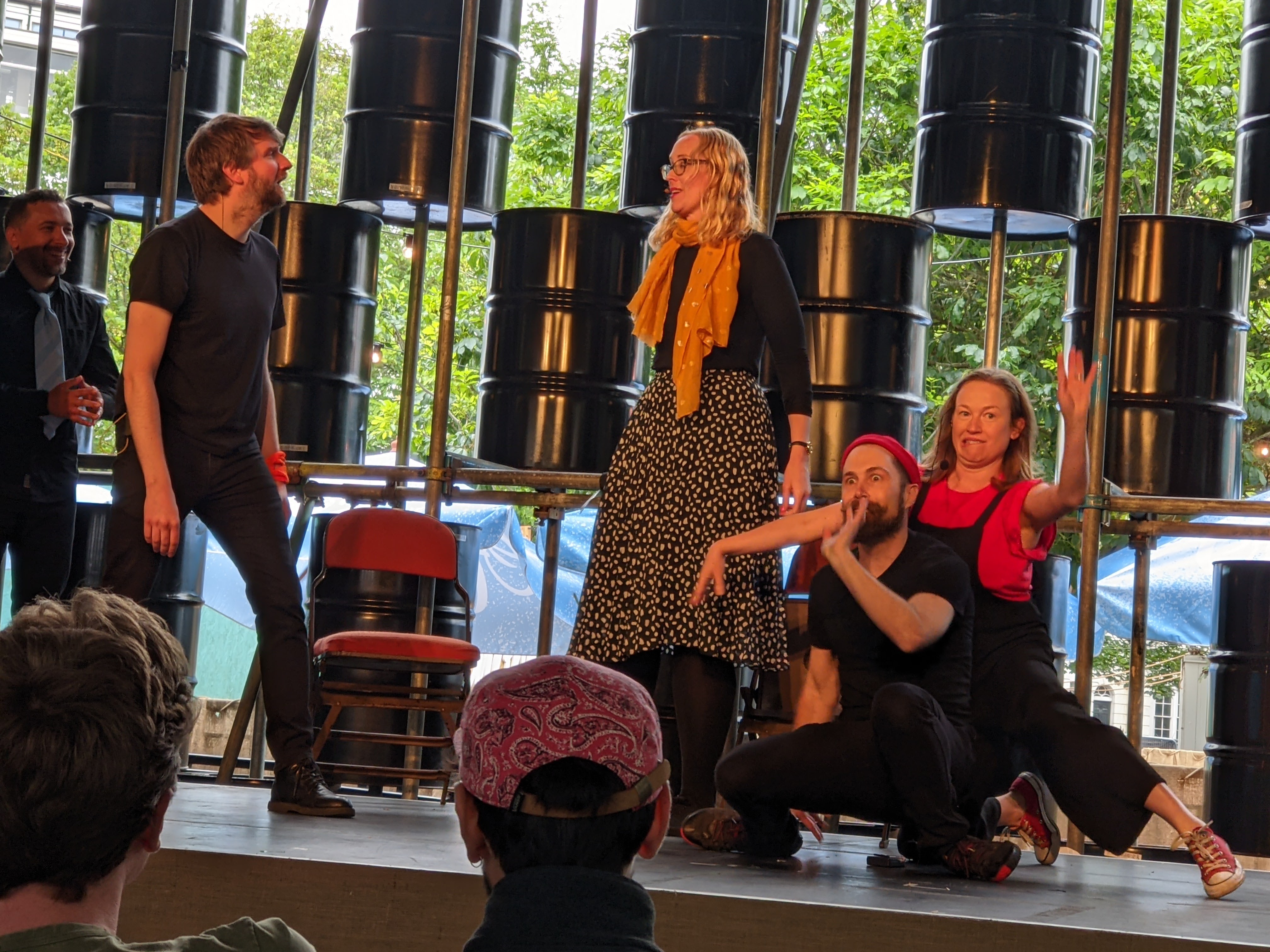}
    \caption{\small{Left: Operator of the AI narrator and virtual avatar. Right: Example of an improvised scene. Photos: Erika Hughes}}
    \label{fig1}
\end{figure}

\subsection{Datasets for AI improv}

Before large language models, conversational agents were trained on datasets geared towards dialogue, like the Cornell Movie Dialogs Corpus \cite{danescu2011chameleons} and OpenSubtitles \cite{tiedemann2009news}. The latter was used to develop conversational models such as \cite{vinyals2015neural} and the improv theatre-specific chatbot A.L.Ex \cite{mathewson2017improvised} from \emph{HumanMachine}\footnote{\small{https://humanmachine.live}}. Later on, an improv-specific dataset of \emph{yes-and} exchanges from improv podcasts was curated in  \cite{cho2020grounding}.


While employed in the context of improvised storytelling, our work departs from generating dialogue and focuses on storytelling from the perspective of a narrator. We hypothesized that the best datasets would come from general fiction novels as well as synopses and plot summaries. Coincidentally, Large Language Models (LLMs) are now pre-trained on comprehensive and diverse sets of corpora and are capable of memorising diverse linguistic patterns from books, novels, movie scripts, newspapers or blog posts \cite{radford2019language,brown2020language}. From the perspective of thematic diversity and specificity, the need for collecting specific training data seems to have become less of an issue. 

\subsection{Live curation, mitigating of model bias}

There are however trade-offs between the predictive power of large language models, and their embedded biases or their misalignment with desired societal values, which have been discussed in \cite{bender2021dangers,kenton2021alignment}.

Our approach to mitigating these biases and to the removal of offensive content relies on a combination of automated filters and human curation, performed in real time in the context of a live show. First, we remove sentences that contain known offensive words from a blocklist, and all generated sentences are validated using multiple filters for inflammatory, hateful or sexual content by the Perspective API\footnote{\small{https://www.perspectiveapi.com/}}. Second, the human who operates the storyteller interface has agency in both how they formulate and type the context, and in what sentences produced by the AI they choose to read, with a possibility to omit or reword parts of those sentences.

\subsection{Interactive live AI narration on stage}

Our interface works in the following way: each time a sentence is typed by the operator, it is concatenated to the context of the scene. GPT-3 is then run $3$ times on the whole context, thus generating three sets of sentences of total length up to $100$ characters for each set. The operator has the choice of selecting none, one, or several of these sentences, in the order they choose\footnote{While it has been observed that GPT-3 is capable of some amount amount of meta-learning, such as recognising and generating analogies, or responding to \emph{commands} (e.g., "translate the following sentence from English to French") \cite{brown2020language}, we decided to limit this work to using GPT-3 as a statistical language model and to leave, for future work, hierarchical text generation or additional prompt engineering, such as ``expand on what happens next'' or ``let's look back at this character''.}.

An important aspect of the human-machine interaction on stage is that the actors' performance and the operation of the AI happen simultaneously, i.e. that the operator types context prompts and chooses AI-generated suggestions at the same time as scenes unfold. The human operator may then interrupt the scene, in a similar way to an improviser `editing` the scene. This delegates to the human cast and to the operator the artistic choices of timing--a crucial element of comedy--and maintains the liveness of the performance.

\subsubsection{Story initiation}

We initially experimented with a system for automated selection of initial writing prompts from the \emph{novel-first-lines-dataset} (a crowdsourced dataset of first sentences of novels)\footnote{\small{https://github.com/janelleshane/novel-first-lines-dataset}}. The single-word audience suggestion would be matched with a fixed set of 11k sentences using sentence-level embeddings computed using the Universal Sentence Encoder \cite{cer2018universal} combined with approximate nearest neighbor search\footnote{\small{https://github.com/korymath/jann}}. Early trials during improv rehearsals demonstrated that the first lines of novels were not informative enough for the actors performing much shorter scenes, and that the actors preferred to initiate the story themselves.

\subsubsection{Avatar for the AI narrator}
We designed a virtual avatar that personified the AI narrator. That avatar consisted of a 3D model of a robot, inspired by Aldebaran Robotics' \emph{Nao}, built using Cinema 4D\footnote{\small{https://www.maxon.net/}} and imported into Adobe Character Animator\footnote{\small{https://www.adobe.com/products/character-animator.html}} as a puppet controlled by facial expressions of the operator as they are reading the AI-generated lines. Instead of using computer-generated voice, we relied on human voice for expressive interpretation. The operator was standing behind the TV screen that was projecting the avatar on the stage.

\subsection{Evaluating AI in performance}
We worked with a small team of professional improvisers to build and rehearse an original $50$ minute performance that included a series of short AI-assisted scenes followed by a $12$-minute AI narrated long-form improvisational scene. 
We then presented $2$ performances for public audiences. We present a partial transcript from one of the AI narrated performances and discuss how well the AI was able to offer contextually relevant suggestions that advanced the plot. 
We also administered anonymous surveys to the performers (p1-p5) as well as $9$ audience members after each performance (a1-a9). The surveys consisted of a series of open ended questions regarding how they experienced the AI on stage. Our surveys were conducted in accordance with the approved ethical standards of our public research institution.

\section{Results}
 
We relate a $12$-minute long-form improvisation between $6$ actors. GPT-3 generated altogether $455$ sentences of suggestions, but only a subset was selected by the operator. The following extract is an exact transcript of only the inputs or \emph{context} given to the GPT-3 language model. {\bf Sentences in bold correspond to GPT-3 outputs, which are fed back again as context}. {\it (Notes in italic and between parentheses are scene descriptions that were not input to GPT-3).} The transcript of performer lines and interactions is not included because of the difficulty of transcribing an improvised performance where many actors may speak on top of one another. The audience gave the suggestion: ``Pizza Hut''.

\begin{displayquote}
At the Pizza Hut. Brian and his date lost patience. {\it (The operator misunderstood the relationship between the two protagonists.)}

There was always a reason for them to admire each other. Brian was an expert at making pizza. Sally found her vocation, making pizza like Brian. Brian started listing all the products... Baguettes, patisserie... Sally asked Brian for help.
{\it (The operator made a confusion in the name, as it was Sandra, not Sally.)}

{\bf The door opened and a burly man entered, followed by his wife.}
{\it (A couple entered the pizzeria, the man spoke with a heavy voice.)}

The husband and the wife entered the pizzeria. They asked for supremes, with garlic bread. 
{\bf Both women had crushes on Brian.}
{\it (The unnamed wife briefly approached Brian.)}

Sally searched for pastries. 
{\bf The husband and the wife asked for vodka.} {\it (Unused suggestion.)}

They got creme patissiere...
{\bf Brian apologized.}
{\it (Sally/Sandra was rolling pizza on the floor.)}

Sally was dreaming about becoming a master patissier.
{\bf She continued to look for pastries.}
{\it (Sally/Sandra said she was done working at Pizza Hut and wanted to resign. Scene transition, with an angry boss entering the stage.)}

Brian's boss told him he would let her go. Sally gave her notice. The boss refused. The boss was cruel.
{\bf Brian asked the boss for her resignation. The boss made a mistake.}
{\it (A confrontation took place between Brian and the boss, the boss later started behaving apologetically.)}
{\bf Brian and Sally left the pizzeria.}
{\it (A male actor stepped in to play the newly introduced Sally.)}
{\it (Scene transition to Sandra at a restaurant owned by the burly man and his wife.)}

Sandra pursued her dream of being a pastry chef. Sandra was serving the old burly couple. 
{\bf The burly man was impressed. The burly man and his wife complimented Sandra.}

Even though Sandra was violating safety regulations. 
{\bf Sandra was getting tired. Sandra's dream would soon come true.}

They loved it! With her sweat, she impressed them. 
{\bf Sandra was now a great pastry chef.}
{\it (Scene transition to the boss joining the group.)}

The boss came to apologise to Sandra. Sandra said that she remembered him. He was diminished. He was wondering if it was safe to do it on the floor... She heard about Brian. Can you come back, he asked. The boss was apologetic. 
{\bf Sandra thanked the boss, who helped her. Brian and Sandra were both happy. Sandra was proud. The boss was really clear. The boss was jealous.}
{\it (Group scene.)}
{\bf He agreed.}
{\it (End scene.)}
\end{displayquote}

As the $12$-minute scene unfolded, the operator was typing a summary of it as inputs to an interface to GPT-3. For each line of context that was input by the operator, there were many alternative suggestions that could have been selected, and this transcript shows only the ones that were actually chosen and presented to the cast and to the audience. The decision to intervene in the narration and the timing and delivery of each intervention were choices made by the operator, who was simultaneously voicing and animating the virtual avatar, as well as observing the live improvised scene.

Just like in the first show (for which we do not report the transcript in this paper), the AI-assisted narrator's interactions became more frequent as the scene was unfolding and the characters established. The motivation for this was to let the actors establish the characters and their relationships first, and to start intervening only once the cast had an initial guess of the narrative arc of the story.

\subsection{Audience and performer response}
We provide the following small sample of $9$ audience responses as useful observations to guide discussion rather than evidence of findings that can be generalised. $7$ of the respondents indicated the presence of AI itself as the most significant motivational factor in attending the events. $6$ reported overall satisfaction with AI narration, $1$ reported neutral satisfaction, and two reported dissatisfaction. The AI narrated scene was the most frequently cited ($5/9$) response to the question `What did you enjoy most about the show.'
 
 All $5$ performers reported satisfaction with the ability of the AI to move the story forward. $3$ however also `slightly agreed' that the AI `mainly introduced absurd or random information' into the scenes. We present the following quotes from performers about their experience to advance discussion about the relationship between the insertion of surprising plot points experienced as both `random' and useful in advancing the story arc. 
 
 \begin{itemize}
     \item  \textit{As a performer I had to physically become the character (the AI)  described in the narration. This pushed me to a certain pov / voice/ physicality which I probably wouldn't have chosen i.e. a gruff, muscly patisserie store owner.} (p1)
     \item \textit{(The AI) added a level of randomness and craziness different from a human brain.} (p2)
     \item \textit{(The AI) really helped the plot move forward, but without being too prescriptive, and enabled me to focus on character development, relationships, emotions and object work.} (p3)
     \item \textit{I did a few scenes as the protagonist where I was sad, and then (the AI) would say 'she was happy' or similar, but I loved that as I has to justify it and it was funny!} (p4)
     \item \textit{Generally the narrative direction (of the AI) helped the show move forward in a good direction} (p5)
 \end{itemize}







\section{Discussion}
In the above exchange between the operator's inputs and the AI suggestions, one can notice that the AI introduced two key characters (the \emph{burly man and his wife}) who played the role of mentors for the main protagonist, \emph{Sandra}, and enabled the resolution of the story by complimenting \emph{Sandra}'s work. The AI's suggestions also satisfied a classical  narrative arc by allowing her ``dream to come true'' and achieving her transformation into a ``great pastry chef''. This illustrates the capacity for an AI-based narrator--operating in tandem with a human curator who makes timing decisions--to generate novel and meaningful plot points.

Interestingly, as (p4) noted, the AI-provided suggestions did not consistently keep the affect or motivational stance of some characters (e.g., the boss was first cruel, then apologetic and even helping Sandra). Where this inconsistency might invalidate a progressing story when uttered from a character (and subsequently fail a Turing-test), in the mouth of a narrator it can encourage performers to maintain classical story arcs that require characters to change and adapt over time \cite{aristotle350poetics}. The inconsistencies of the AI-generated text were interpreted by the cast as narrated reversals of feelings, and challenged the performers (as p1 suggests) to allow themselves to change and be affected by each other in surprising but meaningful ways. In improv theatre this is described as a `status' reversal where the `low' status of a character at the beginning of a scene becomes `high' status by the end \cite{giebel2019improvising}. Such reversals are in practice often difficult for human improvisers to execute, as one instinctively attempts to maintain their given status or fight to maintain `high' status. In this instance, the AI drove the plot more aggressively forward and motivated the performers to shift and adapt status to the evolving circumstance that in effect provided a more clear beginning, middle, and end to the story.

As a creative partner, rather than simply providing strange or absurd plot points to challenge the human performers to make sense out of, the AI seems to have removed some of the cognitive load for improvisers (as with p3) allowing them to concentrate on relationships. Without a narrator, improvisers must both react spontaneously in the moment, and remember to engage narrative techniques such as status changes to move the story forward. This important practice of narrative making can be understood as ``a carefully argued process of removing and adding participants'' \cite{kumar2008algorithms}. The practices of theatrical improvisation and acting techniques such as Meisner actor training \cite{moseley2012meisner} explicitly ask performers ``not to be in their heads'', meaning not to withdraw from the live performance in order to plot or to reflect and comment on the scene, but rather to dedicate their entire attention to what is happening in front of them on the stage. We believe that one of the potential applications of computational creative systems could be to alleviate the cognitive load of performers to shift their focus from plotting to reacting.

Strikingly, the seemingly random characters introduced by the AI were often the result of human error. But even when such errors were introduced the performers playfully accepted the offers that resulted in comic relief, skilfully transforming an error into a serendipitous opportunity to `break the fourth wall' and to connect with the audience. For instance the wrong naming of the main protagonist (as the operator mistakenly and repeatedly typed ``Sally'' instead of ``Sandra'') led the improvises to quickly introduce, then shelve, a temporary character. This tight collaboration between improvisers and AI prevented the introduction of a new character that may have otherwise been considered a `less carefully argued' addition to the narrative, to still perform a useful function (comic relief) without disrupting the evolving story.

\section{Conclusions}

Narrative theatrical performances encapsulate human culture, social interaction, physical expression and natural human emotion. Improv is an ideal test-bed to explore questions about the human-AI collaborative creative capacity. It has been proposed as a \emph{grand challenge} for artificial intelligence \cite{martin2016improvisational}. We believe that AI-as-collaborator, as in this current study, uplifts artists, as opposed to challenging them.


Language models capture statistics of written corpora of human culture, and thus provide human audiences with a mirror of typical narrative tropes and biases. Thus, they highlight the need for human interpretation and curation of AI-generated content. Our two-pronged approach of automated filters followed by human operator selection of sentences, illustrates a transfer of responsibility from the language model to the (human) narrator--not unlike a typical improv show, where the human cast are responsible for the story they tell (e.g., ``punching up, not down'') and adapt to their audiences (e.g., family-friendly vs. late-night shows).


This work is, to the best of our knowledge, the first staging of an AI narrator co-creating improvised theatre alongside humans for a live audience.
Timing and aesthetics are significant factors for the human experience of AI by audiences and cast members.
The ease of use of the narrative interface for the human operator impacts how quickly they can add to the language model context or choose from its outputs. Finally, the imagined `personality' of the AI narrator play a role in co-creation. These are important avenues for future research on human-AI co-creation.



\section{Acknowledgments}
The authors wish to acknowledge the anonymous audience participants who filled the questionnaires, and above all, to thank the actors from improv theatre troupe Improbotics\footnote{\small{https://improbotics.org}} who collectively created the artwork that enabled this specific study, namely Harry Turnbull, Julie Flower, Marouen Mraihi, Paul Little and Sarah Davies.






\bibliographystyle{iccc}
\bibliography{iccc}

\begin{thebibliography}{}

\bibitem[\protect\citeauthoryear{Aristotle}{350 BC}]{aristotle350poetics}
Aristotle.
\newblock 350 B.C.
\newblock {\em Poetics}.
\newblock Translated by S. H. Butcher.

\bibitem[\protect\citeauthoryear{Baumer and Magerko}{2010}]{baumer2010analysis}
Baumer, A., and Magerko, B.
\newblock 2010.
\newblock An analysis of narrative moves in improvisational theatre.
\newblock In {\em JICIDS},  165--175.
\newblock Springer.

\bibitem[\protect\citeauthoryear{Bender \bgroup et al.\egroup
  }{2021}]{bender2021dangers}
Bender, E.~M.; Gebru, T.; McMillan-Major, A.; and Shmitchell, S.
\newblock 2021.
\newblock On the dangers of stochastic parrots: Can language models be too big?
\newblock In {\em Conf. on Fairness, Accountability, and Transparency},
  610--623.

\bibitem[\protect\citeauthoryear{Brown \bgroup et al.\egroup
  }{2020}]{brown2020language}
Brown, T.~B.; Mann, B.; Ryder, N.; Subbiah, M.; Kaplan, J.; Dhariwal, P.;
  Neelakantan, A.; Shyam, P.; Sastry, G.; Askell, A.; et~al.
\newblock 2020.
\newblock Language models are few-shot learners.
\newblock {\em arXiv preprint arXiv:2005.14165}.

\bibitem[\protect\citeauthoryear{Bruce and others}{2000}]{bruce2000robot}
Bruce, A., et~al.
\newblock 2000.
\newblock Robot improv: Using drama to create believable agents.
\newblock In {\em IEEE ICRA}.

\bibitem[\protect\citeauthoryear{Cer \bgroup et al.\egroup
  }{2018}]{cer2018universal}
Cer, D.; Yang, Y.; Kong, S.-y.; Hua, N.; Limtiaco, N.; John, R.~S.; Constant,
  N.; Guajardo-Cespedes, M.; Yuan, S.; Tar, C.; et~al.
\newblock 2018.
\newblock Universal sentence encoder.
\newblock {\em arXiv preprint arXiv:1803.11175}.

\bibitem[\protect\citeauthoryear{Cho and May}{2020}]{cho2020grounding}
Cho, H., and May, J.
\newblock 2020.
\newblock Grounding conversations with improvised dialogues.
\newblock {\em arXiv preprint arXiv:2004.09544}.

\bibitem[\protect\citeauthoryear{Danescu-Niculescu-Mizil and
  Lee}{2011}]{danescu2011chameleons}
Danescu-Niculescu-Mizil, C., and Lee, L.
\newblock 2011.
\newblock Chameleons in imagined conversations: A new approach to understanding
  coordination of linguistic style in dialogs.
\newblock {\em arXiv preprint arXiv:1106.3077}.

\bibitem[\protect\citeauthoryear{Eger and Mathewson}{2018}]{eger2018dairector}
Eger, M., and Mathewson, K.~W.
\newblock 2018.
\newblock dairector: Automatic story beat generation through knowledge
  synthesis.
\newblock {\em arXiv preprint arXiv:1811.03423}.

\bibitem[\protect\citeauthoryear{Giebel}{2019}]{giebel2019improvising}
Giebel, J.~D.
\newblock 2019.
\newblock Improvising tactical choices based on status or “who’s driving
  the dramatic action bus?”.
\newblock {\em Objectives, Obstacles, and Tactics in Practice: Perspectives on
  Activating the Actor}.

\bibitem[\protect\citeauthoryear{Hayes-Roth and
  Van~Gent}{1996}]{hayes1996improvisational}
Hayes-Roth, B., and Van~Gent, R.
\newblock 1996.
\newblock Improvisational puppets, actors, and avatars.
\newblock In {\em Proc Comp Game Dev Conf}.

\bibitem[\protect\citeauthoryear{Jacob}{2019}]{jacob2019improvisational}
Jacob, M.
\newblock 2019.
\newblock {\em Improvisational Artificial Intelligence for Embodied
  Co-creativity}.
\newblock Ph.D. Dissertation, GIT.

\bibitem[\protect\citeauthoryear{Johnstone}{1979}]{johnstone1979}
Johnstone, K.
\newblock 1979.
\newblock {\em Impro: Improvisation and the theatre}.
\newblock London: Faber and Faber Ltd.

\bibitem[\protect\citeauthoryear{Kenton \bgroup et al.\egroup
  }{2021}]{kenton2021alignment}
Kenton, Z.; Everitt, T.; Weidinger, L.; Gabriel, I.; Mikulik, V.; and Irving,
  G.
\newblock 2021.
\newblock Alignment of language agents.
\newblock {\em arXiv preprint arXiv:2103.14659}.

\bibitem[\protect\citeauthoryear{Kumar \bgroup et al.\egroup
  }{2008}]{kumar2008algorithms}
Kumar, D.; Ramakrishnan, N.; Helm, R.~F.; and Potts, M.
\newblock 2008.
\newblock Algorithms for storytelling.
\newblock {\em IEEE Transactions on Knowledge and Data Engineering}
  20(6):736--751.

\bibitem[\protect\citeauthoryear{Magerko and
  others}{2011}]{magerko2011employing}
Magerko, B., et~al.
\newblock 2011.
\newblock Employing fuzzy concept for digital improvisational theatre.
\newblock In {\em AIIDE},  53--60.

\bibitem[\protect\citeauthoryear{Martin, Harrison, and
  Riedl}{2016}]{martin2016improvisational}
Martin, L.~J.; Harrison, B.; and Riedl, M.~O.
\newblock 2016.
\newblock Improvisational computational storytelling in open worlds.
\newblock In {\em International Conference on Interactive Digital
  Storytelling},  73--84.
\newblock Springer.

\bibitem[\protect\citeauthoryear{Mathewson and
  Mirowski}{2017}]{mathewson2017improvised}
Mathewson, K.~W., and Mirowski, P.
\newblock 2017.
\newblock Improvised theatre alongside artificial intelligences.
\newblock In {\em AIIDE}.

\bibitem[\protect\citeauthoryear{Mathewson and
  Mirowski}{2018}]{mathewson2018improbotics}
Mathewson, K.~W., and Mirowski, P.
\newblock 2018.
\newblock Improbotics: Exploring the imitation game using machine intelligence
  in improvised theatre.
\newblock In {\em AAAI AIIDE}.

\bibitem[\protect\citeauthoryear{Nichols, Gao, and
  Gomez}{2020}]{nichols2020collaborative}
Nichols, E.; Gao, L.; and Gomez, R.
\newblock 2020.
\newblock Collaborative storytelling with large-scale neural language models.
\newblock In {\em Motion, Interaction and Games}.
\newblock  1--10.

\bibitem[\protect\citeauthoryear{O’Neill and others}{2011}]{o2011knowledge}
O’Neill, B., et~al.
\newblock 2011.
\newblock A knowledge-based framework for the collaborative improvisation of
  scene introductions.
\newblock In {\em ICIDS},  85--96.
\newblock Springer.

\bibitem[\protect\citeauthoryear{Perlin and Goldberg}{1996}]{perlin1996improv}
Perlin, K., and Goldberg, A.
\newblock 1996.
\newblock Improv: A system for scripting interactive actors in virtual worlds.
\newblock In {\em Conf on Comp. G. \& Int. Tech.}
\newblock ACM.

\bibitem[\protect\citeauthoryear{Radford and
  others}{2019}]{radford2019language}
Radford, A., et~al.
\newblock 2019.
\newblock Language models are unsupervised multitask learners.
\newblock {\em OpenAI Blog} 1(8).

\bibitem[\protect\citeauthoryear{Riedl and Stern}{2006}]{riedl2006believable}
Riedl, M.~O., and Stern, A.
\newblock 2006.
\newblock Believable agents and intelligent story adaptation for interactive
  storytelling.
\newblock In {\em Intl Conf on Tech for Int Dig St and Ent},  1--12.
\newblock Springer.

\bibitem[\protect\citeauthoryear{Spolin and
  Sills}{1963}]{spolin1963improvisation}
Spolin, V., and Sills, P.
\newblock 1963.
\newblock {\em Improvisation for the theater: A handbook of teaching and
  directing techniques}.
\newblock Northwestern University Press.

\bibitem[\protect\citeauthoryear{Tiedemann}{2009}]{tiedemann2009news}
Tiedemann, J.
\newblock 2009.
\newblock News from opus-a collection of multilingual parallel corpora with
  tools and interfaces.
\newblock In {\em Recent Advances in NLP}, volume~5.

\bibitem[\protect\citeauthoryear{Vinyals and Le}{2015}]{vinyals2015neural}
Vinyals, O., and Le, Q.
\newblock 2015.
\newblock A neural conversational model.
\newblock {\em arXiv preprint arXiv:1506.05869}.

\end{thebibliography}

\end{document}